\documentclass[journal=jpccck,manuscript=article,layout=traditional]{achemso}

\usepackage{amsfonts}
\usepackage{amsmath}
\usepackage{amssymb}
\usepackage{graphicx}
\usepackage[]{natbib}
\usepackage{multirow}
\usepackage{dcolumn}%

\usepackage[normalem]{ulem}
\usepackage{epsfig}
\usepackage{float}
\usepackage{amsmath}
\usepackage{soul}
\usepackage[colorinlistoftodos,prependcaption,textsize=tiny]{todonotes} 
\usepackage{xcolor,colortbl}

\definecolor{Gray}{rgb}{0.83,0.83,0.98}
\definecolor{LightCyan1}{rgb}{0.83,0.83,0.98}
\definecolor{LightCyan2}{rgb}{0.91,0.92,1}

\author{Cesar E. P. Villegas}
\affiliation{Istituto di Struttura della Materia of the National Research Council, Via Salaria Km 29.3,I-00016 Monterotondo Stazione, Italy.}
\alsoaffiliation{Instituto de F\'{\i}sica Te\'{o}rica, Universidade Estadual Paulista (UNESP), Rua Dr. Bento T. Ferraz, 271, S\~{a}o Paulo, SP 01140-070, Brazil.}
\email{villegas@ift.unesp.br}
\author{A. R. Rocha}
\affiliation{Instituto de F\'{\i}sica Te\'{o}rica, Universidade Estadual Paulista (UNESP), Rua Dr. Bento T. Ferraz, 271, S\~{a}o Paulo, SP 01140-070, Brazil.}
\author{Andrea Marini}
\affiliation{Istituto di Struttura della Materia of the National Research Council, Via Salaria Km 29.3,I-00016 Monterotondo Stazione, Italy.}
\alsoaffiliation{European Theoretical Spectroscopy Facilities (ETSF)}

\title{Anomalous temperature dependence of the band-gap in Black Phosphorus}

\begin{document}




\begin{abstract}
Black Phosphorus (BP) has gained renewed attention due to its singular anisotropic electronic and optical properties
that might be exploited for a wide range of technological applications. In this respect, the thermal
properties are particularly important both to predict its room temperature operation and to determine its thermoelectric potential.
From this point of view, one of the most spectacular and poorly understood phenomena is, indeed, the BP 
temperature--induced band--gap opening: when temperature is increased the fundamental band--gap increases
instead of decreasing. This anomalous thermal dependence has also been observed, recently, in its monolayer counterpart.
In this work, based on \textit{ab-initio} calculations, we present an explanation for this long known, and yet not fully 
explained, effect. We show that it arises from a combination of harmonic and lattice thermal expansion
contributions, which are, in fact, highly interwined. We clearly narrow down
the mechanisms that cause this gap opening
by identifying the peculiar atomic vibrations that drive the anomaly. The final picture we give explains both the BP anomalous 
band--gap opening and the frequency increase with increasing volume (tension effect).
\end{abstract}
\maketitle
\newline
Recently, black phosphorus (BP), a layered puckered structure of P atoms has gained renewed attention due to its remarkable anisotropic 
optical\cite{aniso2}, electrical\cite{aniso1,rev2} and thermal properties\cite{aniso4}. In contrast to other layered semiconductors, the band gap in BP can 
be adjusted from 0.33 to 2.0 eV depending on the number of layers\cite{tran}. Due to these peculiar properties, it has been rapidly
envisioned the possibility of integrating BP for several technological applications\cite{apli1,apli2}. In fact, recent reports have
indicated that BP and its monolayer counterpart hold great potential for designing thermoelectric devices as their figure of merit
can go from 0.7 to 2.5 under controlled doping and temperature conditions\cite{termo1,termo2}. These findings have highlighted
questions regarding the origin of such high thermoelectric performance and, most importantly, the fundamental role played by temperature. 

The band gap temperature dependence in semiconductors is a well understood phenomena for a large group of materials, for which
one observes a monotonic decrease of the energy gap as temperature increases\cite{cardonarev,Varshni}. Nevertheless, there are 
some exceptional materials that exhibit an anomalous temperature dependence: the gap increases instead of decreasing. 
Two kinds of anomalies are known: non-monotonic and monotonic. In the first case,
the gap increases at low temperatures and decreases for high temperatures\cite{cnt-gap,calcopyri}.
In the second case, instead, the gap continuously increases with temperature. This is the case of some
perovskites\cite{perov1}, Copper halides\cite{cardona2} and lead chalcogenides\cite{tauber,leadchal1}.
Despite the abundance of experimental evidence, and theoretical results based on different approaches\cite{leadchal1,cohen,baleva} and
models,\cite{hauschild,anoma-car} fully \textit{ab-initio} theoretical studies addressing the origin of the anomalous
gap dependence are still scarce.

Some clear experimental evidences\cite{bptemp0,bptemp2}, recently confirmed by the photoluminescence spectrum
of its monolayer counterpart\cite{monoblue}, have indicated that BP follows a monotonic increase of the band-gap. This
trend has been explained neither theoretically nor experimentally. Indeed most vibrational studies performed 
on BP have characterized the phonon modes\cite{Raman1,Raman2} and investigated the phonon thermal transport\cite{aniso4,termo1,termo2}. 
Consequently, the role of the electron-phonon interaction\cite{dresel2} and lattice expansion in black phosphorus are, so far, 
poorly understood.

In this work, we provide a comprehensive description of the anomalous band gap temperature dependence of black
phosphorus. By using \textit{ab-initio} calculations, we study both the harmonic (electron-phonon coupling)
and lattice thermal expansion (variable volume) contributions. For
the harmonic part, the mechanism is driven by the coupling between electronic 
states with acoustic modes and the transverse optical mode $B_{1u}$. The coupling stems from long-range amplitude vibrations that are favored by 
the particular puckered crystal structure. For the lattice thermal expansion (LTE) term the anomaly arises due to the BP negative pressure coefficient.
We also show that BP volume expansion induces a tension effect\cite{nte} via transverse optical vibrations. Our results show an excellent agreement
with experimental $E_{g}(T)$ curves slope and emphasizes the crucial role of both harmonic and LTE contributions.

\begin{figure}
\begin{center}
\includegraphics[width=1.0\columnwidth]{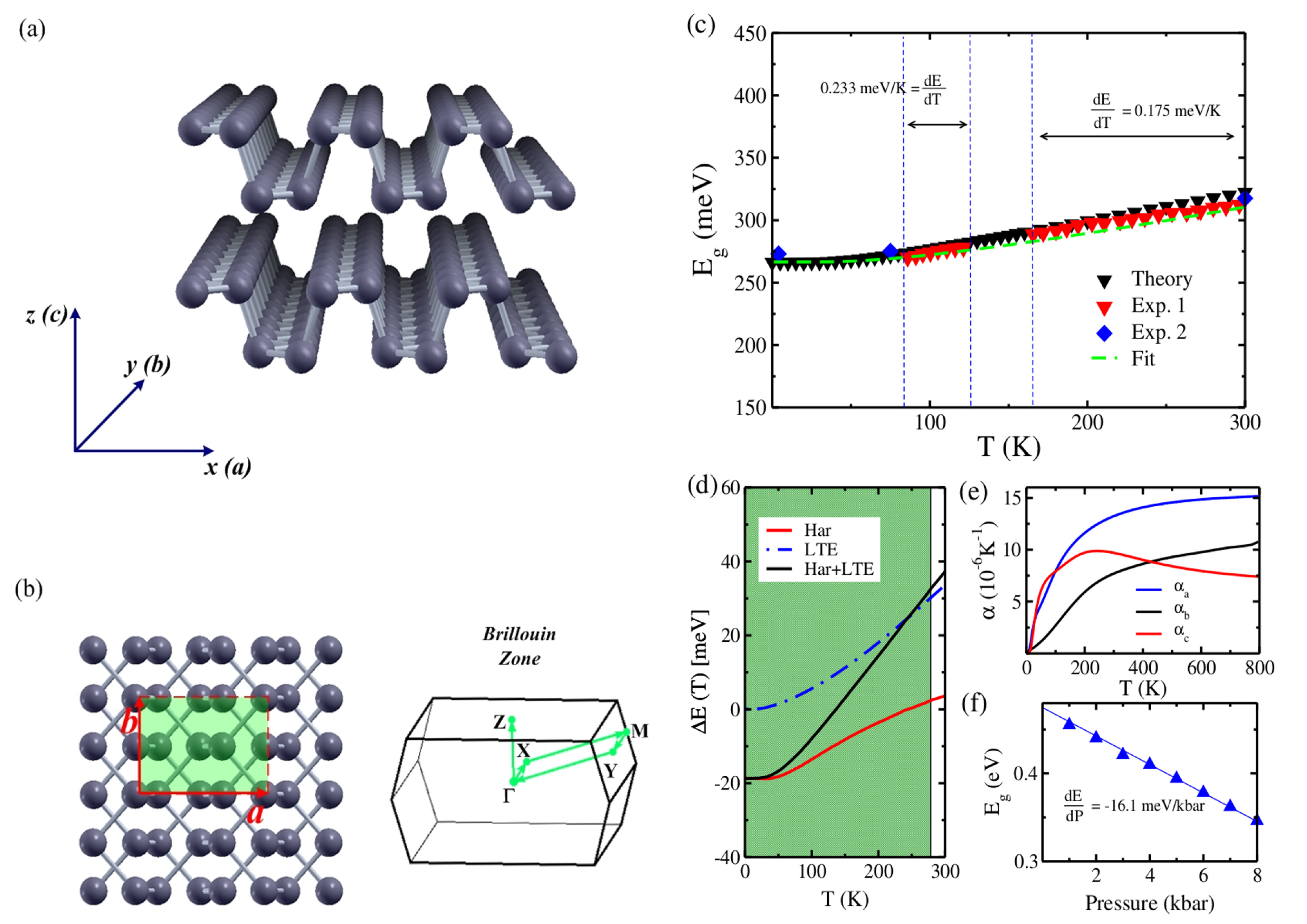}
\caption{\small{ (a-b) Schematic representation of the atomic structure and Brillouin zone of Black Phosphorus showing the 
main crystallographic directions. (c) Theoretical and experimental BP band-gap thermal dependence split in frame (d) in
harmonic and LTE contributions obtained from \textit{ab-initio} MBPT
and the QHA, respectively. The relevant thermodynamic observables required
for computing the LTE term such as (e) the linear expansion coefficients along the different crystal directions and
(f) the energy gap change as a function of pressure, obtained from DFT, are also shown.
Green dashed-line in (c) is a fit with a three oscillator Bose-Einstein model for frequencies 136, 195 and, 450 cm$^{-1}$ which 
represent the most intense peaks in the experimental BP phonon density of states\cite{phdos}. The green shaded region indicates 
the temperature range below 282K.
}}\label{fig1} 
\end{center}
\end{figure}

In the finite temperature regime, the temperature ($T$) dependence of the single particle state energy is 
$E_{n\textbf{k}}(T)=\epsilon_{n\textbf{k}}+\Delta E_{n\textbf{k}}(T)$, with $n\textbf{k}$ the level index and $\epsilon_{n\textbf{k}}$ the energy in the case where all atoms are
kept frozen in their equilibrium position and treated classically. $\Delta E_{n\textbf{k}}(T)$ comprises two contributions
for temperatures below the Debye temperature ($\Theta_{D}$)\cite{foot1}
\begin{equation}
\Delta E_{n\textbf{k}}(T)\approx \Delta E_{n\textbf{k}}(T)\big|_{har} + \Delta E(T)\big|_{LTE}. \label{eq1}
\end{equation}
The first term represents harmonic (constant volume) contributions induced by the pure electron-phonon
interaction. This includes explicitly the temperature dependence through the Bose-Einstein
occupation function of the phonon modes, as explained in the Methods section. The second term describes the LTE (variable volume) effects,
\begin{equation}
 \begin{split}
\Delta E(T)\big|_{LTE}=-\frac{1}{\chi}\frac{\partial E_{g}}{\partial P}\Big|_{T}\int_{0}^{T} \beta(T')dT'. \label{eq1.2}
\end{split}
\end{equation}
In Eq (\ref{eq1.2}), $\beta$ and $\chi$ are the volumetric thermal expansion coefficient and 
compressibility, respectively. Considering the anisotropic crystal structure of BP,
the volumetric coefficient is calculated as $\beta$=$\alpha_{a}$+$\alpha_{b}$+$\alpha_{c}$, where 
$\alpha_{i}$ is the linear expansion coefficient along the crystallographic directions shown in Fig. (\ref{fig1}-a).
Eq. (\ref{eq1}) includes the so-called zero point motion (ZPM) effect\cite{cardonarev,zpm} ($\Delta E_{n\textbf{k}}(T\to0)$)
which is typically non-zero due to the quantum nature of the atoms. 

Moreover, from Eq. (\ref{eq1}) it is clear that the harmonic and LTE terms
must be treated on equal footing. Nevertheless their relative strength strongly depends on the material.
In simple semiconductors with a standard monotonically decreasing gap dependence we have that, generally, $\Delta E(T)\big|_{LTE}$
$\ll$ $\Delta E_{n\textbf{k}}(T)\big|_{har}$.
In systems with an anomalous trend, instead, the role played by the two contributions 
turn out to be of the same order of magnitude\cite{baleva,cohen}. The case of 
PbTe, probably the most well studied lead chalcogenide, is exemplar. PbTe shows a clear band-gap anomaly
of the second kind. This has been interpreted in terms of both strong anharmonic\cite{nature1,science} and 
harmonic\cite{cohen,baleva,leadchal1} effects,
but to date, fully \textit{ab-initio} studies including retardation effects are not available to support this picture.
This example demonstrates that a clear atomistic interpretation of the elemental mechanism that drives the anomaly is still missing. 
In this work, instead, we will unambiguously identify the specific atomic oscillations responsible for
the anomalous gap dependence using a fully \textit{ab-initio} approach.

In our calculations the electron-phonon interaction is treated within the \textit{ab-initio} Many-Body Perturbation Theory (MBPT) 
framework\cite{marini2015,marini2014}. In MBPT the electron-phonon coupling is composed by a first and second order Taylor expansion
in the nuclear displacement, fully including retardation effects caused by the different electronic and phononic
dynamics. The LTE contribution is calculated by using first 
principles within the quasiharmonic approximation (QHA)\cite{qha1}. See Methods section for details.  

In Fig. (\ref{fig1}-c) we show the BP band gap measurements\cite{bptemp2,bptemp0} 
as a function of temperature compared to our theoretical results.
One can clearly observe the monotonic increase of
the experimental band gap nicely described by our calculations. The size of $\Delta E_{g}(T)$--$\Delta E_{g}(T\to0)$ can be as large
as 50 meV at room temperature.
For low-temperatures, we obtain an excellent agreement with the experimental slope while, as the temperature
increases, the theoretical result slightly deviates from the experimental one.
Our interpretation of this deviation is based on the observation that as $T$ increases $\Delta E(T)\big|_{LTE}$ rapidly becomes
dominant. This points to the importance of anharmonic corrections that, however, in Eq. (\ref{eq1}) are not
fully introduced. Thus we ascribe the high temperature deviation to the missing anharmonic corrections to
the electron-phonon and thermal expansion terms.

From the harmonic contribution shown in Fig. (\ref{fig1}-d), we estimate a value of -18.3$\pm$0.9 meV for the ZPM correction.
This is comparable with those reported in other narrow band gap materials\cite{cardonarev}. In addition, we observe that the positive slope 
of the harmonic term smoothly decreases beyond 180K. In contrast, the LTE correction increases continuously reaching 
values as large as 33 mev at room temperature. Note that in Fig. (1-c), the value $\epsilon_{g}$=282.7 meV was added 
to the sum of the harmonic and LTE contributions to match the $E_{g}(T\to0)$ obtained from the fitted
curve using Einstein's model, the green-dashed curve in Fig. (\ref{fig1}-c). This rigid shift does not affect our conclusions as
the main goal of this work is the description of the slope of $E_{g}(T)$, and not the absolute values of the band-gap. This is done in order to avoid the
discrepancies observed in describing the BP band-gap at different theory levels.\cite{tran,qpgap}

The linear expansion and pressure coefficients of BP are shown in Figs (\ref{fig1}-e) and (\ref{fig1}-f). These 
thermodynamic observables, together with the experimental value for the compressibility\cite{compre},
($\chi=3.02 \times 10^{-3}$  kbar$^{-1}$) were used to compute $\Delta E(T)\big|_{LTE}$ in Eq. (\ref{eq1.2}). The
linear expansion coefficients reflect the anisotropic character of the BP crystal as they are different for each
crystallographic direction. The 800K scale is covered in order to show their trend. For completeness,
the lattice parameters as a function of temperature are presented in the Supporting Information. In addition, 
from Fig. (\ref{fig1}-f), we observe that the effect of pressure is to significantly
reduce the band gap of the crystal. By performing a linear fit of the $E_{g}(P)$ data, we estimate a negative pressure coefficient
of $\approx$ -16.1 meV/kbar, which is in excellent agreement with the experimental value of, -16.4 meV/kbar\cite{prescoef,bptemp2}.
It should be noticed that the positive slope observed in the LTE term is a direct consequence of the
negative pressure coefficient as $\Delta E(T)\big|_{LTE} \propto -\frac{\partial E_{g}}{\partial P}$.

The $E_{g}(T)$ slopes for different temperature ranges are summarized in table \ref{table1}. The results show that
the harmonic term is responsible for roughly 50 \% of the total contribution for temperatures up to 160K. However as
the temperature increases (T$>$160K), the electron-phonon contribution gets weaker, with
a slope $\approx$ 0.085 meV/K, indicating that the LTE contribution becomes
the dominant one. 

The correct description of the experimental $E_{g}(T)$ curve is one of the key result of this work. In addition,
the use of an \textit{ab-initio} approach allow us to further investigate the elemental mechanism that drives the BP gap
anomaly. 
\begin{table}[t]
\centering
\caption{Temperature slopes $dE_{g}/dT$ for different temperature ranges obtained by performing a linear fit of $E_g\left(T\right)$.
}\label{table1}

\begin{tabular}{lllll}
\hline
\multicolumn{5}{c}{Temperature Coefficient (meV/K)}    
\\[0.1ex] 
\hline \hline 
\rowcolor{LightCyan1}& Harmonic   & \ \ \ \ LTE & \ \ \ Total & Experiment\cite{bptemp0}
\\[0.5ex]                                       
\rowcolor{LightCyan2}
$\frac{\Delta E_{g}}{\Delta T}|_{80<T<120K}$ & \ \ \ 0.100 &  \ \ \ \  0.99      & \ \ \ \  0.199     &  \ \ \ \ \ \ 0.233       \\[1.5ex]
\rowcolor{LightCyan2}
$\frac{\Delta E_{g}}{\Delta T}|_{160<T<300K}$  & \ \ \ 0.085  & \ \ \ \ 0.150     & \ \ \ \ 0.235      & \ \ \ \ \ 0.175     \\[1.2ex]
\hline \hline

\end{tabular}
\end{table}
\section{Discussion}
\begin{figure}[t]
\begin{center}
\includegraphics[width=1.0\columnwidth]{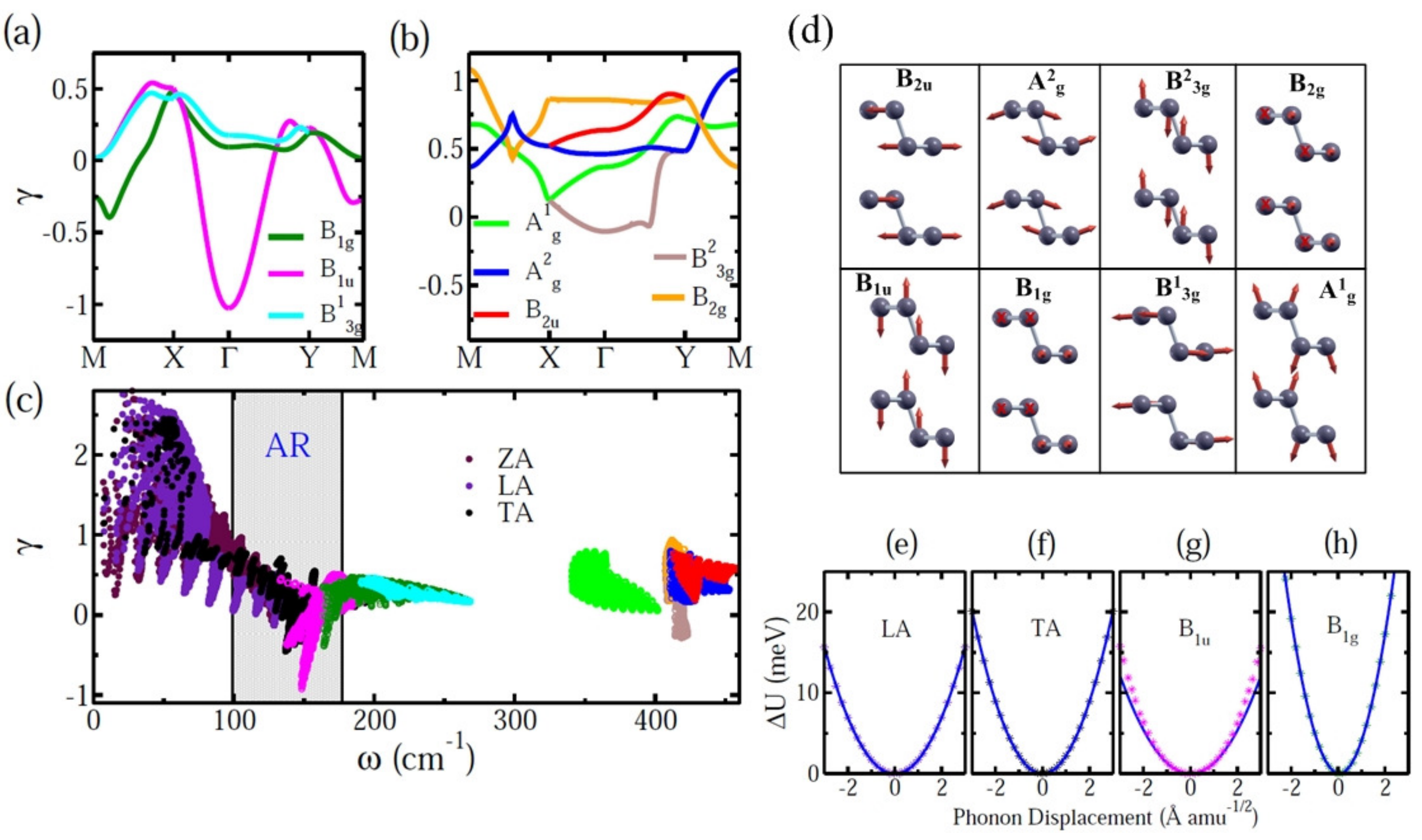}
\caption{\small{ Variable volume (LTE) behaviour of the phonon modes in BP studied within
the QHA. The Gr\"{u}neisen parameters are shown along the $M X \Gamma Y M$ path for: (a) low-frequency and (b) high frequency
optical modes. (c) Gr\"{u}neisen parameters
mapped on the whole Brillouin zone. The anomalous region (AR) is represented with grey-shadow color. (d)
Schematic representation of phonon vibrations in BP. (e-h) Superficial potential energy profile for
LA, TA, $B_{1u}$ and $B_{1g}$ modes. For the acoustic modes the potential is obtained at the M point, whereas the
$\Gamma$ point is used for the optical modes. Solid blue lines represent a parabolic fit.
}} \label{fig2}
\end{center}
\end{figure}

To elucidate the origin of the anomalous thermal behaviour we first analyze the LTE
(volume variable) contribution. For this purpose, we compute the Gr\"{u}neisen parameters
\begin{equation}
 \gamma_{\textbf{q}\lambda}=-\frac{\partial \ln \omega_{\textbf{q}\lambda}}{\partial \ln V}, \label{eq2}
\end{equation}
which allows us to quantify the anharmonicity associated
to a given phonon mode $\lambda$. In Fig. (\ref{fig2}-a) the Gr\"{u}neisen parameters for low-frequency optical modes are presented. We clearly observe
that the $\gamma_{\textbf{q}\lambda}$ of transverse optical (TO) mode $B_{1u}$ shows a strong negative intensity around the $\Gamma$
point, becoming the most anharmonic optical mode at the center of the Brillouin zone (BZ) (see also Fig. \ref{fig2}-b ).
From Eq. (\ref{eq2}), it follows that negative values of $\gamma$ induces an increase of the phonon frequency as the
system's volume expands, contrary to the usual behaviour. This is a well known thermal mechanism, observed also in some vibrational modes
of graphene and graphite\cite{graph}, that lead them to a negative volumetric thermal expansion\cite{foot2}.
Taking a closer view on the vibrational mechanism of mode $B_{1u}$ depicted
in Fig. (\ref{fig2}-d), we argue that this particular negative $\gamma_{\textbf{q}\lambda}$ is a direct consequence of the puckered crystalline 
structure. 
Indeed, during a volumetric expansion, the force constant as well as the frequencies of in-plane bonds are reduced. Simultaneously,
superficial potential energy is gained due to the tension produced among bonds. Thus, the induced in-plane tension will cause a
frequency increase for the atoms vibrating transversally, like in the $B_{1u}$ mode. This is referred in literature as tension effect \cite{nte}
and it is analogous to stretching a string and applying a transversal force which results in higher frequency vibrations. 

In contrast to the transverse optical modes ($B_{1u}$ and $B^{2}_{3g}$), longitudinal
and out-of-plane optical modes shown in Fig. (\ref{fig2}-b),
present positive values for $\gamma_{\lambda}$ at the center of the BZ, reflecting the usual reduction of 
their frequencies as the material expands. Overall, the dispersion
of the Gr\"{u}neisen parameters reflects the structural anisotropy of the material. 

For completeness, in Fig. (\ref{fig2}-c), we present the Gr\"{u}neisen parameter as a function of the phonon frequencies. As expected,
we observe strong anharmonicity in the acoustic modes as they play a crucial role in umklapp scattering
processes. Most importantly, within the frequency range from 100-175 cm$^{-1}$, depicted as a grey-shadow 
region, we observe thermal anomalies (negative value of $\gamma_{\lambda}$) for optical as well as acoustic modes.
As these modes are crucial for both harmonic and LTE terms of 
the anomalous gap dependence, we refer to the highlighted grey region of Fig. (\ref{fig2}-c) as anomalous region (AR).
The $B_{1u}$ mode falls in this range.
In order to understand the vibrational mechanisms in this region, in Figs. (\ref{fig2} e-h) we plot the superficial phonon potential
for modes transverse acoustic (TA), longitudinal acoustic (LA), $B_{1u}$ and $B_{1g}$.
Large phonon amplitudes around the equilibrium
position is observed for modes LA, TA, and $B_{1u}$. This reveals the easiness of these modes
to be scattered when coupled with electrons. In contrast, the small phonon displacements found in
mode $B_{1g}$ remarks it robustness to be scattered. In addition, as the $B_{1u}$ phonon amplitude increases,
anharmonic features appear. This is observed by fitting an harmonic potential (parabola), indicated by the blue-solid line
in Fig. (\ref{fig2} e-h).

The link between the negative Gr\"{u}neisen parameters, the anomalous gap dependence and a family of specific modes may appear
fortuitous. Instead these modes, belonging to the AR, are the only driving force of the entire and peculiar thermal behaviour of BP.
\begin{figure} 
\begin{center}
\includegraphics[width=0.6\columnwidth]{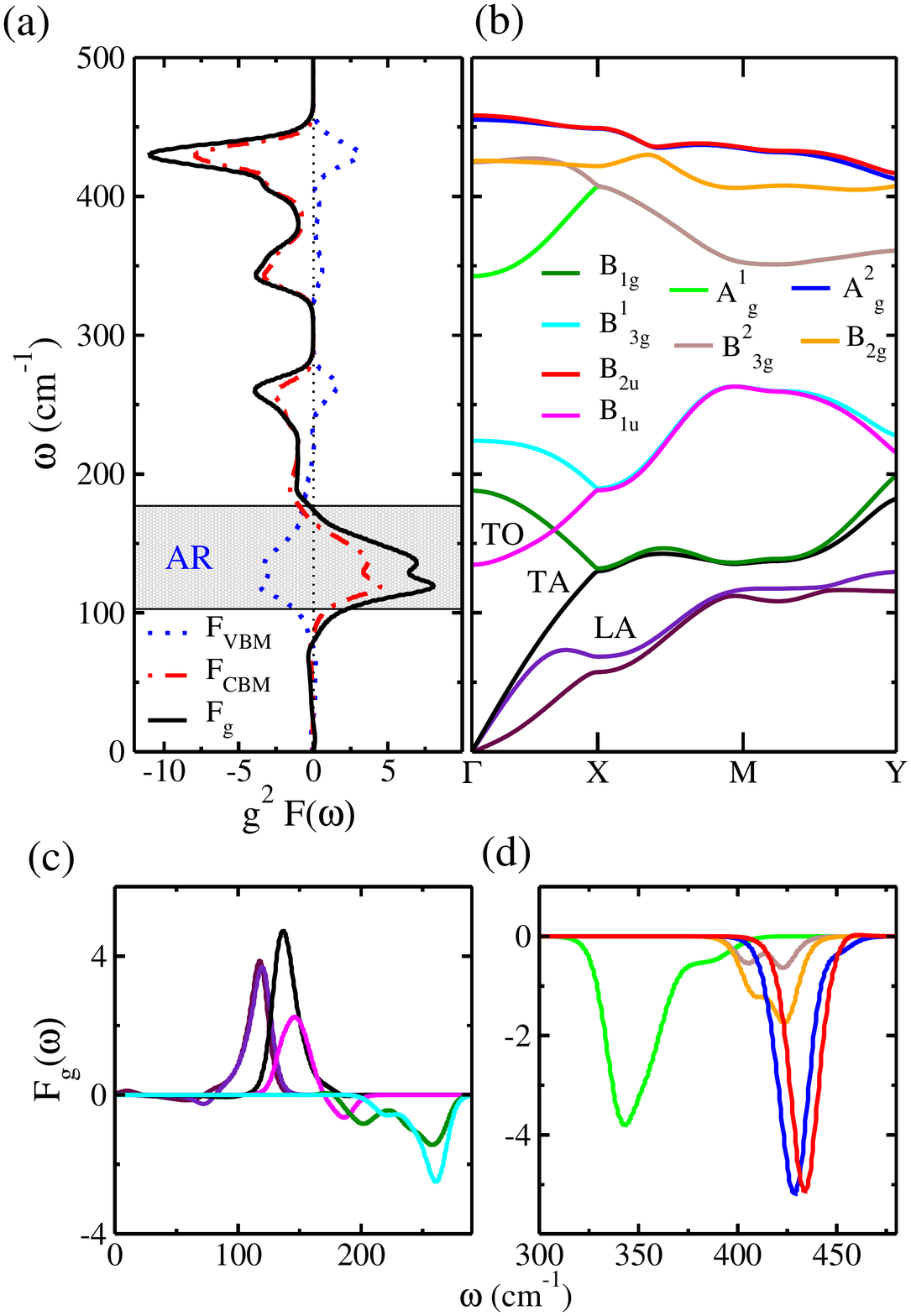}
\caption{\small{Constant volume (harmonic) behaviour of BP phonon modes studied within MBPT. (a) Electron-phonon spectral function 
for VBM (blue dotted-line), CBM (red dot-dashed line) and band-edge (black solid line).
As $\Delta E_{g}(T)\big|_{har}\propto \int d\omega g^{2}F_{g}(\omega)[N_{\textbf{q}\lambda}(\omega,T)+1/2]$, the positive $g^{2}F_{g}$ region
corresponds to a temperature increasing $E_{g}(T)$. As in the LTE part
the modes of the AR drive the anomaly (gray-shadow region). (b) BP phonon dispersion along the $\Gamma X M Y$ path and band edge Eliashberg spectral function projected on
phonon modes with frequencies (c) bellow 300 cm$^{-1}$ and (d) above 300 cm$^{-1}$.
}} \label{fig3}
\end{center}
\end{figure}

Indeed, in order to investigate the phonon mode contributions to 
$\Delta E_{g}(T)\big|_{har}$, we introduce the generalized Eliashberg function
\begin{equation}
 \begin{split}
g^{2}F_{n\textbf{k}}(\omega)=
\sum_{n^{'}\textbf{q}\lambda }\Big[\frac{|g^{\textbf{q}\lambda}_{nn'\textbf{k}}|^2}{\epsilon_{n\textbf{k}}
-\epsilon_{n'\textbf{k-q}}}-\frac{1}{2}\frac{\Lambda^{\textbf{q}\lambda}_{nn'\textbf{k}}}
{\epsilon_{n\textbf{k}}-\epsilon_{n'\textbf{k}}}\Big]
\delta(\omega-\omega_{\textbf{q}\lambda}). \label{eq3}
 \end{split}
\end{equation}

This function is defined within the adiabatic limit which corresponds to the condition $|\epsilon_{n\textbf{k}}-\epsilon_{n'\textbf{k-q}}| \gg \omega_{\textbf{q}\lambda}$. 
In this work we use the fully dynamical description for $\Delta E_{n\textbf{k}}(T)\big|_{har}$ (see Methods section). Nevertheless, the adiabatic $g^{2}F_{n\textbf{k}}(\omega)$ 
provides useful insights into the decomposition and physical understanding of the temperature dependence of the electronic states
as it defines the adiabatic contribution to $\Delta E_{n\textbf{k}}(T)\big|_{har}$ as
\begin{equation}
\Delta E_{n\textbf{k}}(T)\big|_{har}^{adiab} \approx \int d\omega g^{2}F_{n\textbf{k}}(\omega)[N_{\textbf{q}\lambda}(\omega,T)+\frac{1}{2}]. \label{eq3.1}
\end{equation}
Dynamical corrections are defined in Eq. (\ref{eq6}). In Eq. (\ref{eq3.1}), $N_{\textbf{q}\lambda}(\omega,T)$ is the Bose-Einstein distribution while in Eq. (\ref{eq3}), 
$\omega_{\textbf{q}\lambda}$ are the phonon frequencies and $g^{\textbf{q}\lambda}_{nn'\textbf{k}}$ and $\Lambda^{\textbf{q}\lambda}_{nn'\textbf{k}}$ are the first and second-order electron-phonon matrix elements,
which are related to the Fan and Debye-Waller (DW) contributions, respectively (see Methods section). 

The sign of the spectral function is determined by the denominator of Eq. (\ref{eq3}), which is connected
to all possible scattering mechanisms involving the emission or absorption of phonons.
Typically, for conventional semiconductors, the Eliashberg function is positive at the valence band maximum (VBM)
and negative at the conduction band minimum (CBM)\cite{cardona3,cardona4}. These opposite trends lead to the
usual reduction of the band gap with temperature. Therefore, they provide a clear tool to investigate the
anomalies induced by the electron-phonon interaction. 

In Fig. (\ref{fig3}-a), we show the Eliashberg function for the CBM and VBM states. It behaves
as in conventional semiconductors for high frequency optical modes. However, in the AR they 
present opposite trends. By defining the band-edge Eliashberg function, $F_{g}$=$F_{CBM}$--$F_{VBM}$,
we see that two very intense positive peaks appear in the AR. They are responsible for the positive slope in $E_{gap}(T)$
as one can inferred directly from Eq. (\ref{eq3.1}) by identifying the positive sign of the integral. Indeed, the
AR is located in the low energy part of the phonon spectrum and, therefore, it is always weighted by the Bose distribution
more than all other modes. Thus, we have the confirmation that the same phonon modes are responsible
for both the harmonic and anharmonic anomalous gap correction. 

The BP phonon dispersion shown in Fig. (\ref{fig3}-b) allow us to further inspect the origin of
the $g^{2}F_{n}$ peaks. A careful analysis of the band-edge $g^{2}F_{n}$ projected 
on each vibrational mode, shown in Fig. (\ref{fig3}-c), clearly indicates that the peak around 146 $cm^{-1}$ 
arises from the TA and TO modes. TA providing the most intense contribution.
It should be noticed that the TA and TO modes have been previously\cite{anoma-car} identified as the main sources for the anomalous
band-gap temperature dependence (with TA being the predominant) in CuBr and AgGaS$_{2}$. This is consistent
with our fully \textit{ab-initio} results based on MBPT. 

As we already observed in the LTE contribution analysis, the electron-phonon coupling related to the anomaly
arises due to large phonon vibration amplitudes around the equilibrium position that are favored by the particular crystal structure of BP.
Note that the mode $B_{1g}$ does not contribute in the origin of the anomaly as its electron-phonon coupling
is very weak in the anomalous region. These results confirm the predictions based on the superficial potential 
plots shown in Fig. (\ref{fig2}-h).
For completeness, in Fig. (\ref{fig3}-d), we show $g^{2}F_{n}$ for high frequency optical modes.
From there we note that peak around 342 cm$^{-1}$ stems from the coupling with the 
mode $A^{1}_{g}$. In addition, the origin of the high frequency peak is traced back and related mainly to
the infrared active $B_{2u}$ and the Raman active $A^{2}_{g}$ modes.
This results are in agreement with a recent work in which the electron-phonon coupling in BP is assessed by means of
group theory.\cite{dresel2}

In summary, by means of \textit{ab-initio} calculations, we have shown that black phosphorus presents 
an anomalous band gap temperature dependence that arises due to electron-phonon coupling with acoustic and
low-frequency transverse optical mode that also cause a negative pressure coefficient. The nature of the electron-phonon coupling is
traced back to the large-amplitude atomic vibrations which are favored by the particular BP layered structure. Our results
spotlight the importance of both harmonic and lattice thermal expansion
contributions on determining the band-gap thermal dependence. We also put in evidence the direct correlation 
between harmonic and anharmonic effects that can be used as fingerprints to identify thermal anomalies in other semiconductors. 


\section{Methods}
The harmonic (constant volume) contribution to the temperature dependence of the electronic states is
described within the Many-Body perturbation theory (MBPT) framework. In MBPT, the electron-phonon interaction is 
treated perturbatively and composed by a first and a second order Taylor expansion in the nuclear displacement\cite{marini2015,marini2014}, which 
are commonly known as the Fan and Debye-Waller (DB) terms, respectively. The interacting Green's function, whose poles define
the quasiparticle excitations, can be written as
\begin{equation}
G_{n\textbf{k}}(\omega,T)=[\omega-\epsilon_{n\textbf{k}}-\Sigma^{Fan}_{n\textbf{k}}(\omega,T)-\Sigma^{DW}_{n\textbf{k}}(T)]^{-1},
\end{equation}
where $\epsilon_{n\textbf{k}}$ is the ground-state frozen atoms eigenenergies, obtained from plane-wave DFT\cite{pwscf}.
$\Sigma^{Fan}$ is the Fan contribution\cite{elena-tesis} 
\begin{equation}
 \begin{split}
\Sigma^{Fan}_{n\textbf{k}}(i\omega,T)=\sum_{n'\textbf{q} \lambda}\frac{|g^{\textbf{q}\lambda}_{nn'\textbf{k}}|^2}{N}\Big[\frac{N_{
 \textbf{q}\lambda}(T)+1-f_{n'\textbf{k-q}}}{i\omega-\epsilon_{n'\textbf{k-q}}-\omega_{\textbf{q}\lambda}}
+\frac{N_{\textbf{q}\lambda}(T)+f_{n'\textbf{k-q}}}{i\omega-\epsilon_{n'\textbf{k-q}}+\omega_{\textbf{q}\lambda}}\Big], \label{eq4}
 \end{split}
\end{equation}
and $\Sigma^{DW}$ is the Debye-Waller term
\begin{equation}
 \Sigma^{DW}_{n\textbf{k}}(T)=-\frac{1}{2}\sum_{n'\textbf{q}\lambda}\frac{\Lambda^{\textbf{q}\lambda}_{nn'\textbf{k}}}{N}
  \Big[\frac{2N_{\textbf{q}\lambda}(T)+1}{\epsilon_{n\textbf{k}}-\epsilon_{n'\textbf{k}}}\Big]. \label{eq5}
\end{equation}
Here $N_{\textbf{q}\lambda}$, $f_{n'\textbf{k-q}}$ represent the Bose-Einstein and Fermi-Dirac distribution functions, while
$\omega_{\textbf{q}\lambda}$ and $N$ are the phonon frequencies and number of $\textbf{q}$ points taken randomly in
the Brillouin zone.
The electron-phonon matrix elements $g^{\textbf{q}\lambda}_{nn'\textbf{k}}$, which represent the probability amplitude
for an electron to be scattered with emission or absorption of phonons is given by\cite{q-random}
\begin{equation}
g^{\textbf{q}\lambda}_{nn'\textbf{k}}=\sum_{s \alpha}[2M_{s}\omega_{\textbf{q} \lambda}]^{-1/2}e^{i\textbf{q}.\tau_{s}} \\
\langle n'\textbf{k-q}|\frac{\partial V_{scf}(\textbf{r})}{\partial R_{s \alpha}}|n\textbf{k} \rangle \xi_{\alpha}(\textbf{q} \lambda|s), \label{eq5.1}
\end{equation}
where, $M_{s}$ is the atomic mass, $\tau_{s}$ is the position of the atomic displacement in the unit cell
and $\xi_{\alpha}(\textbf{q}\lambda|s)$ are the components of the phonon polarization vectors. $V_{scf}(r)$ is the self-consistent
DFT ionic potential. $\Lambda^{\textbf{q}\lambda}_{nn'\textbf{k}}$ are the second-order electron-phonon matrix elements
which were calculated within the rigid-ion approximation.\cite{q-random,marini2014} A detailed discussion and derivation of the electron-phonon matrix elements,
including $\Lambda^{\textbf{q}\lambda}_{nn'\textbf{k}}$ can be found elsewhere\cite{marini2014,marini2015}.

By assuming the quasiparticle approximation (QPA) to expand in first-order the self-energy frequency dependence
around the bare energies, one can write the non-adiabatic temperature dependent quasiparticle energy corrections as\cite{Marini20091392}
\begin{equation}
\Delta E_{n\textbf{k}}(T)\big|_{har}=E_{n\textbf{k}}(T)-\epsilon_{n\textbf{k}} \approx Z_{n\textbf{k}}(T)[\Sigma_{n\textbf{k}}^{Fan}(\epsilon_{n\textbf{k}},T)+\Sigma_{n\textbf{k}}^{DW}(T)],
 \label {eq6}
\end{equation}
with the renormalization factor $Z_{n\textbf{k}}(T)$=$\big[1$-$\frac{\partial \Re \Sigma_{n\textbf{k}}^{Fan}(\omega)}{\partial \omega}\big|_{\omega=\epsilon_{n\textbf{k}}}\big]^{-1}$.
Given that the Fan self-energy term is a complex function, it provides both an energy renormalization shift 
and an intrinsic quasiparticle lifetime. It is worth to mention that the MBPT formalism presented here represents
the dynamical extension of the Heine-Allen-Cardona approach\cite{cardona3} which is naturally recovered from eq. ({\ref{eq4}}) by assuming
$|\epsilon_{n\textbf{k}}-\epsilon_{n'\textbf{k-q}}| \gg \omega_{\textbf{q}\lambda}$ (the \textit{adiabatic} limit) and 
$\omega \approx \epsilon_{n\textbf{k}}$ (the \textit{on-the-mass-shell} limit). Thus, the adiabatic corrections can be written as
\begin{equation}
\Delta E_{n\textbf{k}}(T)\big|_{har}^{adiab} \approx \Sigma_{n\textbf{k}}^{Fan}(\epsilon_{n\textbf{k}},T)+\Sigma_{n\textbf{k}}^{DW}(T)=
\int d\omega g^{2}F_{n\textbf{k}}(\omega)[N_{\textbf{q}\lambda}(\omega,T)+\frac{1}{2}].
 \label {eq6.1}
\end{equation}
The ground-state charge density and electronic states, necessary to evaluated Eq. (\ref{eq6}), were obtained by performing plane-wave
density functional theory (DFT) using the Perdew-Burke-Ernzerhof (PBE) exchange-correlation functional currently implemented in the
Quantum-Espresso code\cite{pwscf}. Norm-conserving pseudopotentials with 3s3p core-states were adopted. The Brillouin zone was
mapped using 6$\times$8$\times$3 k-sampling with 80 Ry kinetic energy cut-off. The electron-phonon matrix
elements necessary to compute the Fan and Debye-Waller self-energies, Eq. (\ref{eq4}) and (\ref{eq5}), were obtained within density
functional perturbation theory (DFPT). In addition, the self-energies were evaluated with 400 electronic bands and 240 random 
\textbf{q}-points for the phonon momentum. These parameters were chosen after a careful convergence tests on the spectral function.
The curves showing the convergence as a function of the electronic bands and \textbf{q}-points can be found in the
supporting information.

The phonon frequencies dependence on volume (anharmonic contribution) is studied by means of the quasiharmonic approximation. 
Within this formalism, the equilibrium volume of a crystal for any temperature $T$, at a constant pressure, can be evaluated by minimizing 
the Helmholtz free energy function with respect the volume\cite{qha,qha1},
\begin{equation}
 F(\{V_{i}\},T)=E(\{V_{i}\})+\sum\limits_{\textbf{q}\lambda}\frac{\hbar \omega_{\textbf{q}\lambda}(\{V_{i}\})}{2}+k_{B}T\sum\limits_{\textbf{q}\lambda}
 \ln(1-\exp[-\frac{\hbar \omega_{\textbf{q}\lambda}(\{V_{i}\})}{k_{B}T}]),
\end{equation}
where the function $E(V_{i})$ is the ground-state energy, obtained from DFT calculations by applying uniaxial strain that varies by $\pm$ 2\%
at each crystallographic direction. Once the Helmholtz function is evaluated at different volumes, the data is fitted with the third-order
Birch-Murnaghan equation of state. 
The linear thermal expansion coefficients of the cell are calculated as
\begin{equation}
\alpha_{i}(T)=\frac{1}{a_{i}(T)}\frac{da_{i}(T)}{dT}.
\end{equation} 
For the anharmonic contributions and energy gap pressure dependence, Van der Waals corrections were taken into account within the
vdW-DF approach for the exchange-correlation functional. This was done in order to open the electronic gap and avoid the
metal transition of the structure during the volume change. The anharmonic properties of the system were 
computed using the Phonopy package\cite{phonopy}. To this aim, a 2$\times$2$\times$2 supercell was constructed.
Evaluating the Gr\"{u}neisen parameters required to remove the system from the equilibrium volume, by 
changing the the lattice parameter by $\pm$ 0.35 \% 
\begin{acknowledgement}
CEPV acknowledges the financial support from the Brazilian agency FAPESP grant number 2015/14899-0 and 2012/24227-1.
AM acknowledges the funding received from the Futuro in Ricerca grant No. RBFR12SW0J of the Italian Ministry of 
Education, University and Research MIUR, the European Union project MaX Materials design at the eXascale
H2020-EINFRA-2015-1, Grant agreement n. 676598 and Nanoscience Foundries and Fine Analysis - Europe 
H2020-INFRAIA-2014-2015, Grant agreement n. 654360. A. R. R. acknowledges support from ICTP-SAIRF (FAPESP project 2011/11973-4) 
and the ICTP-Simons Foundation Associate Scheme. This work uses the computational resources from  GRID-UNESP and CENAPAD/SP.
\end{acknowledgement}

\begin{suppinfo}
Convergence test related to the temperature dependent electronic states;
lattice constant as a function of temperature; lattice 
thermal expansion as a function of T. 
\end{suppinfo}


\begin{mcitethebibliography}{50}
\providecommand*\natexlab[1]{#1}
\providecommand*\mciteSetBstSublistMode[1]{}
\providecommand*\mciteSetBstMaxWidthForm[2]{}
\providecommand*\mciteBstWouldAddEndPuncttrue
  {\def\EndOfBibitem{\unskip.}}
\providecommand*\mciteBstWouldAddEndPunctfalse
  {\let\EndOfBibitem\relax}
\providecommand*\mciteSetBstMidEndSepPunct[3]{}
\providecommand*\mciteSetBstSublistLabelBeginEnd[3]{}
\providecommand*\EndOfBibitem{}
\mciteSetBstSublistMode{f}
\mciteSetBstMaxWidthForm{subitem}{(\alph{mcitesubitemcount})}
\mciteSetBstSublistLabelBeginEnd
  {\mcitemaxwidthsubitemform\space}
  {\relax}
  {\relax}

\bibitem[Wang \latin{et~al.}(2015)Wang, Jones, Tran, Jia, Zhao, Wang, Yang, Xu,
  and Xia]{aniso2}
Wang,~X.; Jones,~A.~M.; Tran,~K. L. S.~V.; Jia,~Y.; Zhao,~H.; Wang,~H.;
  Yang,~L.; Xu,~X.; Xia,~F. \emph{Nat. Nanotech.} \textbf{2015}, \emph{10},
  517--521\relax
\mciteBstWouldAddEndPuncttrue
\mciteSetBstMidEndSepPunct{\mcitedefaultmidpunct}
{\mcitedefaultendpunct}{\mcitedefaultseppunct}\relax
\EndOfBibitem
\bibitem[Xia \latin{et~al.}(2015)Xia, Wang, and Jia]{aniso1}
Xia,~F.; Wang,~H.; Jia,~Y. \emph{Nat. Commun.} \textbf{2015}, \emph{5},
  4458\relax
\mciteBstWouldAddEndPuncttrue
\mciteSetBstMidEndSepPunct{\mcitedefaultmidpunct}
{\mcitedefaultendpunct}{\mcitedefaultseppunct}\relax
\EndOfBibitem
\bibitem[Ling \latin{et~al.}(2015)Ling, Wang, Huang, Xia, and
  Dresselhaus]{rev2}
Ling,~X.; Wang,~H.; Huang,~S.; Xia,~F.; Dresselhaus,~M.~S. \emph{Proc. Natl.
  Acad. Sci} \textbf{2015}, \emph{112}, 201416581\relax
\mciteBstWouldAddEndPuncttrue
\mciteSetBstMidEndSepPunct{\mcitedefaultmidpunct}
{\mcitedefaultendpunct}{\mcitedefaultseppunct}\relax
\EndOfBibitem
\bibitem[Luo \latin{et~al.}(2015)Luo, Maassen, Deng, Du, Garrelts, Lundstrom,
  Ye, and Xu]{aniso4}
Luo,~Z.; Maassen,~J.; Deng,~Y.; Du,~Y.; Garrelts,~R.~P.; Lundstrom,~M.~S.;
  Ye,~P.~D.; Xu,~X. \emph{Nat. Commun.} \textbf{2015}, \emph{6}, 8572\relax
\mciteBstWouldAddEndPuncttrue
\mciteSetBstMidEndSepPunct{\mcitedefaultmidpunct}
{\mcitedefaultendpunct}{\mcitedefaultseppunct}\relax
\EndOfBibitem
\bibitem[Tran \latin{et~al.}(2014)Tran, Soklaski, Liang, and Yang]{tran}
Tran,~V.; Soklaski,~R.; Liang,~Y.; Yang,~L. \emph{Phys. Rev. B} \textbf{2014},
  \emph{89}, 235319\relax
\mciteBstWouldAddEndPuncttrue
\mciteSetBstMidEndSepPunct{\mcitedefaultmidpunct}
{\mcitedefaultendpunct}{\mcitedefaultseppunct}\relax
\EndOfBibitem
\bibitem[Li \latin{et~al.}(2014)Li, Yu, Ye, Ge, Ou, Wu, Feng, Chen, and
  Zhang]{apli1}
Li,~L.; Yu,~Y.; Ye,~G.~J.; Ge,~Q.; Ou,~X.; Wu,~H.; Feng,~D.; Chen,~X.~H.;
  Zhang,~Y. \emph{Nat. Nanotech.} \textbf{2014}, \emph{9}, 372--377\relax
\mciteBstWouldAddEndPuncttrue
\mciteSetBstMidEndSepPunct{\mcitedefaultmidpunct}
{\mcitedefaultendpunct}{\mcitedefaultseppunct}\relax
\EndOfBibitem
\bibitem[Yuan \latin{et~al.}(2015)Yuan, Liu, Afshinmanesh, Li, Xu, Sun, Lian,
  Curto, Ye, Hikita, Shen, Zhang, Chen, Brongersma, Hwang, and Cui]{apli2}
Yuan,~H. \latin{et~al.}  \emph{Nat. Nanotech.} \textbf{2015}, \emph{10},
  707--713\relax
\mciteBstWouldAddEndPuncttrue
\mciteSetBstMidEndSepPunct{\mcitedefaultmidpunct}
{\mcitedefaultendpunct}{\mcitedefaultseppunct}\relax
\EndOfBibitem
\bibitem[Qin \latin{et~al.}(2014)Qin, Yan, Qin, Yue, Cui, Zheng, and
  Su]{termo1}
Qin,~G.; Yan,~Q.-B.; Qin,~Z.; Yue,~S.-Y.; Cui,~H.-J.; Zheng,~Q.-R.; Su,~G.
  \emph{Sci. Rep.} \textbf{2014}, \emph{4}, 6949\relax
\mciteBstWouldAddEndPuncttrue
\mciteSetBstMidEndSepPunct{\mcitedefaultmidpunct}
{\mcitedefaultendpunct}{\mcitedefaultseppunct}\relax
\EndOfBibitem
\bibitem[Fei \latin{et~al.}(2014)Fei, Faghaninia, Soklaski, Yan, Lo, and
  Yang]{termo2}
Fei,~R.; Faghaninia,~A.; Soklaski,~R.; Yan,~J.-A.; Lo,~C.; Yang,~L. \emph{Nano
  Lett.} \textbf{2014}, \emph{14}, 6393--6399\relax
\mciteBstWouldAddEndPuncttrue
\mciteSetBstMidEndSepPunct{\mcitedefaultmidpunct}
{\mcitedefaultendpunct}{\mcitedefaultseppunct}\relax
\EndOfBibitem
\bibitem[Cardona and Thewalt(2005)Cardona, and Thewalt]{cardonarev}
Cardona,~M.; Thewalt,~M. L.~W. \emph{Rev. Mod. Phys.} \textbf{2005}, \emph{77},
  1173\relax
\mciteBstWouldAddEndPuncttrue
\mciteSetBstMidEndSepPunct{\mcitedefaultmidpunct}
{\mcitedefaultendpunct}{\mcitedefaultseppunct}\relax
\EndOfBibitem
\bibitem[Varshni(1967)]{Varshni}
Varshni,~Y.~P. \emph{Physica} \textbf{1967}, \emph{34}, 149--154\relax
\mciteBstWouldAddEndPuncttrue
\mciteSetBstMidEndSepPunct{\mcitedefaultmidpunct}
{\mcitedefaultendpunct}{\mcitedefaultseppunct}\relax
\EndOfBibitem
\bibitem[Lefebvre \latin{et~al.}(2004)Lefebvre, Finnie, and Homma]{cnt-gap}
Lefebvre,~J.; Finnie,~P.; Homma,~Y. \emph{Phys. Rev. B} \textbf{2004},
  \emph{70}, 045419\relax
\mciteBstWouldAddEndPuncttrue
\mciteSetBstMidEndSepPunct{\mcitedefaultmidpunct}
{\mcitedefaultendpunct}{\mcitedefaultseppunct}\relax
\EndOfBibitem
\bibitem[Bhosale \latin{et~al.}(2012)Bhosale, Ramdas, Burger, Mu{\~n}oz,
  Romero, Cardona, Lauck, and Kremer]{calcopyri}
Bhosale,~J.; Ramdas,~A.~K.; Burger,~A.; Mu{\~n}oz,~A.; Romero,~A.~H.;
  Cardona,~M.; Lauck,~R.; Kremer,~R.~K. \emph{Phys. Rev. B} \textbf{2012},
  \emph{86}, 195208\relax
\mciteBstWouldAddEndPuncttrue
\mciteSetBstMidEndSepPunct{\mcitedefaultmidpunct}
{\mcitedefaultendpunct}{\mcitedefaultseppunct}\relax
\EndOfBibitem
\bibitem[Yu \latin{et~al.}(2011)Yu, Chen, Wang, Pfenninger, Vockic, Kenney, and
  Shum]{perov1}
Yu,~C.; Chen,~Z.; Wang,~J.~J.; Pfenninger,~W.; Vockic,~N.; Kenney,~J.~T.;
  Shum,~K. \emph{J. Appl. Phys.} \textbf{2011}, \emph{110}, 063526\relax
\mciteBstWouldAddEndPuncttrue
\mciteSetBstMidEndSepPunct{\mcitedefaultmidpunct}
{\mcitedefaultendpunct}{\mcitedefaultseppunct}\relax
\EndOfBibitem
\bibitem[Gobel \latin{et~al.}(1998)Gobel, Ruf, Cardona, Lin, Wrzesinski,
  Steube, Reimann, Merle, and Joucla]{cardona2}
Gobel,~A.; Ruf,~T.; Cardona,~M.; Lin,~C.~T.; Wrzesinski,~J.; Steube,~M.;
  Reimann,~K.; Merle,~J.-C.; Joucla,~M. \emph{Phys. Rev. B} \textbf{1998},
  \emph{57}, 15183\relax
\mciteBstWouldAddEndPuncttrue
\mciteSetBstMidEndSepPunct{\mcitedefaultmidpunct}
{\mcitedefaultendpunct}{\mcitedefaultseppunct}\relax
\EndOfBibitem
\bibitem[Tauber \latin{et~al.}(1966)Tauber, Machonis, and Cadoff]{tauber}
Tauber,~R.~N.; Machonis,~A.~A.; Cadoff,~I.~B. \emph{Jour. Appl. Phys.}
  \textbf{1966}, \emph{37}, 4855\relax
\mciteBstWouldAddEndPuncttrue
\mciteSetBstMidEndSepPunct{\mcitedefaultmidpunct}
{\mcitedefaultendpunct}{\mcitedefaultseppunct}\relax
\EndOfBibitem
\bibitem[Gibbs \latin{et~al.}(2013)Gibbs, Kim, Wang, White, Kaviany, and
  Snyder]{leadchal1}
Gibbs,~Z.~M.; Kim,~H.; Wang,~H.; White,~R.~L.; Kaviany,~F. D.~M.; Snyder,~G.~J.
  \emph{Appl. Phys. Lett.} \textbf{2013}, \emph{103}, 262109\relax
\mciteBstWouldAddEndPuncttrue
\mciteSetBstMidEndSepPunct{\mcitedefaultmidpunct}
{\mcitedefaultendpunct}{\mcitedefaultseppunct}\relax
\EndOfBibitem
\bibitem[Tsang and Cohen(1971)Tsang, and Cohen]{cohen}
Tsang,~Y.~W.; Cohen,~M.~L. \emph{Phys. Rev. B} \textbf{1971}, \emph{3},
  1254\relax
\mciteBstWouldAddEndPuncttrue
\mciteSetBstMidEndSepPunct{\mcitedefaultmidpunct}
{\mcitedefaultendpunct}{\mcitedefaultseppunct}\relax
\EndOfBibitem
\bibitem[Baleva \latin{et~al.}(1990)Baleva, Georgiev, and Lashkarev]{baleva}
Baleva,~M.; Georgiev,~T.; Lashkarev,~G. \emph{J. Phys.: Condens. Matter}
  \textbf{1990}, \emph{1}, 2935--2940\relax
\mciteBstWouldAddEndPuncttrue
\mciteSetBstMidEndSepPunct{\mcitedefaultmidpunct}
{\mcitedefaultendpunct}{\mcitedefaultseppunct}\relax
\EndOfBibitem
\bibitem[Hauschild \latin{et~al.}(2006)Hauschild, Priller, Decker,
  Br{\"u}ckner, Kalt, and Klingshirn]{hauschild}
Hauschild,~R.; Priller,~H.; Decker,~M.; Br{\"u}ckner,~J.; Kalt,~H.;
  Klingshirn,~C. \emph{phys. stat. sol. (c)} \textbf{2006}, \emph{3},
  976--979\relax
\mciteBstWouldAddEndPuncttrue
\mciteSetBstMidEndSepPunct{\mcitedefaultmidpunct}
{\mcitedefaultendpunct}{\mcitedefaultseppunct}\relax
\EndOfBibitem
\bibitem[Cardona and Kremer(2014)Cardona, and Kremer]{anoma-car}
Cardona,~M.; Kremer,~R.~K. \emph{Thin Solid Films} \textbf{2014}, \emph{571},
  680--683\relax
\mciteBstWouldAddEndPuncttrue
\mciteSetBstMidEndSepPunct{\mcitedefaultmidpunct}
{\mcitedefaultendpunct}{\mcitedefaultseppunct}\relax
\EndOfBibitem
\bibitem[Baba \latin{et~al.}(1991)Baba, Nakamura, Shibata, and Morita]{bptemp0}
Baba,~M.; Nakamura,~Y.; Shibata,~K.; Morita,~A. \emph{Jpn. J. Appl. Phys.}
  \textbf{1991}, \emph{30}, L1178\relax
\mciteBstWouldAddEndPuncttrue
\mciteSetBstMidEndSepPunct{\mcitedefaultmidpunct}
{\mcitedefaultendpunct}{\mcitedefaultseppunct}\relax
\EndOfBibitem
\bibitem[Morita(1986)]{bptemp2}
Morita,~A. \emph{Appl. Phys. A} \textbf{1986}, \emph{39}, 227--242\relax
\mciteBstWouldAddEndPuncttrue
\mciteSetBstMidEndSepPunct{\mcitedefaultmidpunct}
{\mcitedefaultendpunct}{\mcitedefaultseppunct}\relax
\EndOfBibitem
\bibitem[Surrente \latin{et~al.}(2016)Surrente, Mitioglu, Galkowski, Tabis,
  Maude, and Plochocka]{monoblue}
Surrente,~A.; Mitioglu,~A.~A.; Galkowski,~K.; Tabis,~W.; Maude,~D.~K.;
  Plochocka,~P. \emph{Phys. Rev. B} \textbf{2016}, \emph{93}, 121405(R)\relax
\mciteBstWouldAddEndPuncttrue
\mciteSetBstMidEndSepPunct{\mcitedefaultmidpunct}
{\mcitedefaultendpunct}{\mcitedefaultseppunct}\relax
\EndOfBibitem
\bibitem[Ribeiro \latin{et~al.}(2016)Ribeiro, Villegas, Bahamon, Muraca,
  Castro-Neto, de~Souza, Rocha, Pimenta, and de~Matos]{Raman1}
Ribeiro,~H.~B.; Villegas,~C. E.~P.; Bahamon,~D.~A.; Muraca,~D.;
  Castro-Neto,~A.~H.; de~Souza,~E. A.~T.; Rocha,~A.~R.; Pimenta,~M.~A.;
  de~Matos,~C. J.~S. \emph{Nat. Comm.} \textbf{2016}, \emph{7}, 12191\relax
\mciteBstWouldAddEndPuncttrue
\mciteSetBstMidEndSepPunct{\mcitedefaultmidpunct}
{\mcitedefaultendpunct}{\mcitedefaultseppunct}\relax
\EndOfBibitem
\bibitem[Ling \latin{et~al.}(2015)Ling, Liang, Huang, Puretzky, Geohegan,
  Sumpter, Kong, Meunier, and Dresselhaus]{Raman2}
Ling,~X.; Liang,~L.; Huang,~S.; Puretzky,~A.~A.; Geohegan,~D.~B.;
  Sumpter,~B.~G.; Kong,~J.; Meunier,~V.; Dresselhaus,~M.~S. \emph{Nano Lett.}
  \textbf{2015}, \emph{15}, 4080--4088\relax
\mciteBstWouldAddEndPuncttrue
\mciteSetBstMidEndSepPunct{\mcitedefaultmidpunct}
{\mcitedefaultendpunct}{\mcitedefaultseppunct}\relax
\EndOfBibitem
\bibitem[Ling \latin{et~al.}(2016)Ling, Huang, Hasdeo, Liang, Parkin, Tatsumi,
  Nugraha, Puretzky, Das, Sumpter, Geohegan, Kong, Saito, Drndic, Meunier, and
  Dresselhaus]{dresel2}
Ling,~X. \latin{et~al.}  \emph{Nano Lett.} \textbf{2016}, \emph{16},
  2260--2267\relax
\mciteBstWouldAddEndPuncttrue
\mciteSetBstMidEndSepPunct{\mcitedefaultmidpunct}
{\mcitedefaultendpunct}{\mcitedefaultseppunct}\relax
\EndOfBibitem
\bibitem[Barrera \latin{et~al.}(2005)Barrera, Bruno, Barron, and Allan]{nte}
Barrera,~G.~D.; Bruno,~J. A.~O.; Barron,~T. H.~K.; Allan,~N.~L. \emph{J. Phys.
  Condens. Matter} \textbf{2005}, \emph{17}, R127--R252\relax
\mciteBstWouldAddEndPuncttrue
\mciteSetBstMidEndSepPunct{\mcitedefaultmidpunct}
{\mcitedefaultendpunct}{\mcitedefaultseppunct}\relax
\EndOfBibitem
\bibitem[foo()]{foot1}
$\Delta E_{n}\big|_{har}$ and $\Delta E_{n}\big|_{LTE}$ are approximated as
  both neglect higher order corrections. Here we assume that, as far as the
  electron-phonon channel is concerned, the Migdal's theorem holds and we can
  safely stick to the harmonic description. For the lattice expansion term,
  based on the quasi harmonic approximation, its accuracy is assumed for
  temperatures bellow $\Theta_{D}$=$h\nu_{D}/k_{B}$ = 281.72K. Here $\nu_{D}$=
  5.85 THz, was estimated as the value of the highest frequency normal
  mode.\relax
\mciteBstWouldAddEndPunctfalse
\mciteSetBstMidEndSepPunct{\mcitedefaultmidpunct}
{}{\mcitedefaultseppunct}\relax
\EndOfBibitem
\bibitem[Cannuccia and Marini(2011)Cannuccia, and Marini]{zpm}
Cannuccia,~E.; Marini,~A. \emph{Phys. Rev. Lett.} \textbf{2011}, \emph{107},
  255501\relax
\mciteBstWouldAddEndPuncttrue
\mciteSetBstMidEndSepPunct{\mcitedefaultmidpunct}
{\mcitedefaultendpunct}{\mcitedefaultseppunct}\relax
\EndOfBibitem
\bibitem[Delaire \latin{et~al.}(2011)Delaire, Ma, Marty, May, McGuire, Du,
  Singh, Podlesnyak, Ehlers, Lumsden, and Sales]{nature1}
Delaire,~O.; Ma,~J.; Marty,~K.; May,~A.~F.; McGuire,~M.~A.; Du,~M.-H.;
  Singh,~D.~J.; Podlesnyak,~A.; Ehlers,~G.; Lumsden,~M.~D.; Sales,~B.~C.
  \emph{Nat. Mater.} \textbf{2011}, \emph{10}, 614\relax
\mciteBstWouldAddEndPuncttrue
\mciteSetBstMidEndSepPunct{\mcitedefaultmidpunct}
{\mcitedefaultendpunct}{\mcitedefaultseppunct}\relax
\EndOfBibitem
\bibitem[Bo\v{z}in \latin{et~al.}(2010)Bo\v{z}in, Malliakas, Souvatzis,
  Proffen, Spaldin, Kanatzidis, and Billinge]{science}
Bo\v{z}in,~E.~S.; Malliakas,~C.~D.; Souvatzis,~P.; Proffen,~T.; Spaldin,~N.~A.;
  Kanatzidis,~M.~G.; Billinge,~S. J.~L. \emph{Science} \textbf{2010},
  \emph{630}, 1660--1663\relax
\mciteBstWouldAddEndPuncttrue
\mciteSetBstMidEndSepPunct{\mcitedefaultmidpunct}
{\mcitedefaultendpunct}{\mcitedefaultseppunct}\relax
\EndOfBibitem
\bibitem[Marini \latin{et~al.}(2015)Marini, Ponc\'e, and Gonze]{marini2015}
Marini,~A.; Ponc\'e,~S.; Gonze,~X. \emph{Phys. Rev. B.} \textbf{2015},
  \emph{91}, 224310\relax
\mciteBstWouldAddEndPuncttrue
\mciteSetBstMidEndSepPunct{\mcitedefaultmidpunct}
{\mcitedefaultendpunct}{\mcitedefaultseppunct}\relax
\EndOfBibitem
\bibitem[Ponc\'e \latin{et~al.}(2014)Ponc\'e, Antonius, Gillet, Boulanger,
  Janssen, Marini, C\^ot\'e, and Gonze]{marini2014}
Ponc\'e,~S.; Antonius,~G.; Gillet,~Y.; Boulanger,~P.; Janssen,~J.~L.;
  Marini,~A.; C\^ot\'e,~M.; Gonze,~X. \emph{Phys. Rev. B.} \textbf{2014},
  \emph{90}, 214304\relax
\mciteBstWouldAddEndPuncttrue
\mciteSetBstMidEndSepPunct{\mcitedefaultmidpunct}
{\mcitedefaultendpunct}{\mcitedefaultseppunct}\relax
\EndOfBibitem
\bibitem[Dove(1993)]{qha1}
Dove,~M.~T. \emph{Introduction to lattice dynamics}; Cambridge university
  press, 1993\relax
\mciteBstWouldAddEndPuncttrue
\mciteSetBstMidEndSepPunct{\mcitedefaultmidpunct}
{\mcitedefaultendpunct}{\mcitedefaultseppunct}\relax
\EndOfBibitem
\bibitem[Kaneta \latin{et~al.}(1986)Kaneta, Yoshida, and Morita]{phdos}
Kaneta,~C.; Yoshida,~H.~K.; Morita,~A. \emph{J. Phys. Soc. Jpn.} \textbf{1986},
  \emph{55}, 1213--1223\relax
\mciteBstWouldAddEndPuncttrue
\mciteSetBstMidEndSepPunct{\mcitedefaultmidpunct}
{\mcitedefaultendpunct}{\mcitedefaultseppunct}\relax
\EndOfBibitem
\bibitem[Rudenko and Katsnelson(2014)Rudenko, and Katsnelson]{qpgap}
Rudenko,~A.~N.; Katsnelson,~M.~I. \emph{Phys. Rev. B} \textbf{2014}, \emph{89},
  201408(R)\relax
\mciteBstWouldAddEndPuncttrue
\mciteSetBstMidEndSepPunct{\mcitedefaultmidpunct}
{\mcitedefaultendpunct}{\mcitedefaultseppunct}\relax
\EndOfBibitem
\bibitem[Riedner \latin{et~al.}(1974)Riedner, Srinivasa, Cartz, Worlton,
  Klinger, and Beyerlein]{compre}
Riedner,~R.~J.; Srinivasa,~S.~R.; Cartz,~L.; Worlton,~T.~G.; Klinger,~R.;
  Beyerlein,~R. \emph{AIP Conf. Proc} \textbf{1974}, \emph{17}, 8\relax
\mciteBstWouldAddEndPuncttrue
\mciteSetBstMidEndSepPunct{\mcitedefaultmidpunct}
{\mcitedefaultendpunct}{\mcitedefaultseppunct}\relax
\EndOfBibitem
\bibitem[Akahama and Kawamura(2001)Akahama, and Kawamura]{prescoef}
Akahama,~Y.; Kawamura,~H. \emph{physica status solidi (b)} \textbf{2001},
  \emph{223}, 349--353\relax
\mciteBstWouldAddEndPuncttrue
\mciteSetBstMidEndSepPunct{\mcitedefaultmidpunct}
{\mcitedefaultendpunct}{\mcitedefaultseppunct}\relax
\EndOfBibitem
\bibitem[Mounet and Marzari(2005)Mounet, and Marzari]{graph}
Mounet,~N.; Marzari,~N. \emph{Phys. Rev. B} \textbf{2005}, \emph{71},
  205214\relax
\mciteBstWouldAddEndPuncttrue
\mciteSetBstMidEndSepPunct{\mcitedefaultmidpunct}
{\mcitedefaultendpunct}{\mcitedefaultseppunct}\relax
\EndOfBibitem
\bibitem[foo()]{foot2}
The Gr{\"u}neisen parameter is linked to other thermodynamic observables
  through the relation $\gamma= \beta V/C_{V} \chi$, where $C_{V}$ is the heat
  capacity and $\gamma=\sum_{\lambda} \gamma_{\lambda}
  C_{\lambda}/\sum_{\lambda} C_{\lambda}$, is defined as the weighted partial
  contribution for each mode $\lambda$. Similarly, $C_{\lambda}$ is the heat
  capacity contribution of each phonon mode. Given that $\chi$, $V$, $C_{V}$
  $>$ 0 then, negative $\gamma$ implies $\beta$ $<$ 0 which means a particular
  material should contracts as the temperature increases.\relax
\mciteBstWouldAddEndPunctfalse
\mciteSetBstMidEndSepPunct{\mcitedefaultmidpunct}
{}{\mcitedefaultseppunct}\relax
\EndOfBibitem
\bibitem[Allen and Cardona(1983)Allen, and Cardona]{cardona3}
Allen,~P.~B.; Cardona,~M. \emph{Phys. Rev. B} \textbf{1983}, \emph{27},
  4760\relax
\mciteBstWouldAddEndPuncttrue
\mciteSetBstMidEndSepPunct{\mcitedefaultmidpunct}
{\mcitedefaultendpunct}{\mcitedefaultseppunct}\relax
\EndOfBibitem
\bibitem[Lautenschlager \latin{et~al.}(1985)Lautenschlager, Allen, and
  Cardona]{cardona4}
Lautenschlager,~P.; Allen,~P.~B.; Cardona,~M. \emph{Phys. Rev. B}
  \textbf{1985}, \emph{31}, 2163\relax
\mciteBstWouldAddEndPuncttrue
\mciteSetBstMidEndSepPunct{\mcitedefaultmidpunct}
{\mcitedefaultendpunct}{\mcitedefaultseppunct}\relax
\EndOfBibitem
\bibitem[Giannozzi \latin{et~al.}(2009)Giannozzi, Baroni, Bonini, Calandra,
  Car, Cavazzoni, Ceresoli, Cococcioni, Dabo, Corso, de~Gironcoli, Fabris,
  Fratesi, Gebauer, Gougoussis, Kokalj, Lazzeri, Martin-Samos, Marzari, Mauri,
  Mazzarello, Paolini, Pasquarello, Paulatto, Sbraccia, Scandolo, Sclauzero,
  Seitsonen, Smogunov, Umari, and Wentzcovitch]{pwscf}
Giannozzi,~P. \latin{et~al.}  \emph{J. Phys. Condens. Matter} \textbf{2009},
  \emph{21}, 395502\relax
\mciteBstWouldAddEndPuncttrue
\mciteSetBstMidEndSepPunct{\mcitedefaultmidpunct}
{\mcitedefaultendpunct}{\mcitedefaultseppunct}\relax
\EndOfBibitem
\bibitem[Cannuccia(2011)]{elena-tesis}
Cannuccia,~E. \emph{Giant Polaronic Effects in Polymers: Breakdown of the
  Quasiparticle Picture}; PhD thesis, Rome Tor Vergata University, 2011\relax
\mciteBstWouldAddEndPuncttrue
\mciteSetBstMidEndSepPunct{\mcitedefaultmidpunct}
{\mcitedefaultendpunct}{\mcitedefaultseppunct}\relax
\EndOfBibitem
\bibitem[Ponc\'e \latin{et~al.}(2014)Ponc\'e, Antonius, Boulanger, Cannuccia,
  Marini, C\^ot\'e, and Gonze]{q-random}
Ponc\'e,~S.; Antonius,~G.; Boulanger,~P.; Cannuccia,~E.; Marini,~A.;
  C\^ot\'e,~M.; Gonze,~X. \emph{Comp. Mat. Sc.} \textbf{2014}, \emph{83},
  341--348\relax
\mciteBstWouldAddEndPuncttrue
\mciteSetBstMidEndSepPunct{\mcitedefaultmidpunct}
{\mcitedefaultendpunct}{\mcitedefaultseppunct}\relax
\EndOfBibitem
\bibitem[Marini \latin{et~al.}(2009)Marini, Hogan, Gr{\"u}ning, and
  Varsano]{Marini20091392}
Marini,~A.; Hogan,~C.; Gr{\"u}ning,~M.; Varsano,~D. \emph{Comp. Phys. Commun.}
  \textbf{2009}, \emph{180}, 1392\relax
\mciteBstWouldAddEndPuncttrue
\mciteSetBstMidEndSepPunct{\mcitedefaultmidpunct}
{\mcitedefaultendpunct}{\mcitedefaultseppunct}\relax
\EndOfBibitem
\bibitem[Togo \latin{et~al.}(2010)Togo, Chaput, Tanaka, and Hug]{qha}
Togo,~A.; Chaput,~L.; Tanaka,~I.; Hug,~G. \emph{Phys. Rev. B} \textbf{2010},
  \emph{81}, 174301\relax
\mciteBstWouldAddEndPuncttrue
\mciteSetBstMidEndSepPunct{\mcitedefaultmidpunct}
{\mcitedefaultendpunct}{\mcitedefaultseppunct}\relax
\EndOfBibitem
\bibitem[Togo and Tanaka(2015)Togo, and Tanaka]{phonopy}
Togo,~A.; Tanaka,~I. \emph{Scr. Mater.} \textbf{2015}, \emph{108}, 1--5\relax
\mciteBstWouldAddEndPuncttrue
\mciteSetBstMidEndSepPunct{\mcitedefaultmidpunct}
{\mcitedefaultendpunct}{\mcitedefaultseppunct}\relax
\EndOfBibitem
\end{mcitethebibliography}
\providecommand{\latin}[1]{#1}
\providecommand*\mcitethebibliography{\thebibliography}
\csname @ifundefined\endcsname{endmcitethebibliography}
  {\let\endmcitethebibliography\endthebibliography}{}

\end{document}